\documentstyle[jkas]{article}

\beginpage{1}
\endpage{4}
\year{2013}\volume{00}\month{March}

\runningauthor {S. Trippe} 
\runningtitle{Derivation of MOND} 

\month{March} \year{2013} \volume{00} \issueno{00}
\beginpage{1}\endpage{4}
\date{Received 2013 February 14; Revised 2013 March 22; Accepted 2013 March 28}

\begin{document}

\def\Mo{{M_{0}}}
\def\Me{{M_{\rm ex}}}
\def\Mt{{M_{\rm tot}}}
\def\am{{a_{\sc m}}}
\def\ao{{a_0}}
\def\ac{{a_{\rm c}}}
\def\gn{{g_{\sc n}}}
\def\vc{{v_{\rm c}}}
\def\vo{{v_{0}}}
\def\al{{\langle}}
\def\ar{{\rangle}}
\def\deg{{^{\circ}}}

\title{A DERIVATION OF MODIFIED NEWTONIAN DYNAMICS}

\author{Sascha Trippe\vspace{0.5mm}}

\address{Department of Physics and Astronomy, Seoul National University, Seoul 151-742, South Korea\\ {\it E-mail : trippe@astro.snu.ac.kr}}

\address{\normalsize{\it (Received 2013 February 14; Revised 2013 March 22; Accepted 2013 March 28)}}

\abstract{\noindent Modified Newtonian Dynamics (MOND) is a possible solution for the missing mass problem in galactic dynamics; its predictions are in good agreement with observations in the limit of weak accelerations. However, MOND does not derive from a physical mechanism and does not make predictions on the transitional regime from Newtonian to modified dynamics; rather, empirical transition functions have to be constructed from the boundary conditions and comparison to observations. I compare the formalism of classical MOND to the scaling law derived from a toy model of gravity based on virtual massive gravitons (the ``graviton picture'') which I proposed recently. I conclude that MOND naturally derives from the ``graviton picture'' at least for the case of non-relativistic, highly symmetric dynamical systems. This suggests that -- to first order -- the ``graviton picture'' indeed provides a valid candidate for the physical mechanism behind MOND and gravity on galactic scales in general.}


\maketitle

\section{INTRODUCTION \label{sect_intro}}

\begin{quote}
\small
{\bf ``}\,It is worth remembering that all of the discussion [on dark matter] so far has been based on the premise that Newtonian gravity and general relativity are correct on large scales. In fact, there is little or no direct evidence that conventional theories of gravity are correct on scales much larger than a parsec or so.\,{\bf ''}

--- \citet{binney1987}, Ch.\,10.4, p.\,635
\end{quote}

\noindent
Since the seminal works by \citet{zwicky1933} and \cite{rubin1980} it has become evident (e.g., \citealt{binney1987,sanders1990}) that the dynamical masses of galaxies and galaxy clusters exceed their luminous (baryonic) masses by up to one order of magnitude -- the well-known \emph{missing mass problem}. A possible solution is provided by \emph{Modified Newtonian Dynamics} (MOND) which postulates a modification of the classical Newtonian laws of inertia and/or gravity in the limit of weak accelerations \citep{milgrom1983a,milgrom1983b,milgrom1983c,bekenstein1984,sanders2002,bekenstein2006,ferreira2009,famaey2012}. Assuming a test particle on a circular orbit around a baryonic mass $\Mo$ at distance $r$ with circular speed\footnote{For simplicity, I only regard absolute values of velocities and accelerations; the orientations are evident from the assumed geometry.} $\vc$, MOND relates the centripetal acceleration $\ac=\vc^2/r$ and the Newtonian acceleration $\gn=G\Mo/r^2$, with $G$ being Newton's constant, like

\begin{equation}
\label{eq_mu}
\frac{\gn}{\ac} = \mu(x)
\end{equation}

\noindent
where $x=\ac/\am$, $\am\approx10^{-10}$\,m\,s$^{-2}$ is \emph{Milgrom's constant}, and $\mu(x)$ is a \emph{transition function} with the asymptotic behavior $\mu(x)\rightarrow1$ for $x\gg1$ and $\mu(x)\rightarrow x$ for $x\ll1$. The first limiting case corresponds to standard Newtonian dynamics. The second limiting case leads to $\vc^4 \approx G\,\Mo\,\am=const.$; this explains the asymptotic flattening of galactic rotation curves, the Tully-Fisher/Faber-Jackson relations, and (via division by $r^2$) the surface brightness--acceleration relation of galaxies (see, e.g., \citealt{famaey2012} for a recent review). 

Despite its success, MOND is obviously incomplete. First, it does not derive from a physical mechanism a priori. Second, even though it provides the correct limiting cases by construction, MOND does not provide $\mu(x)$ itself and makes no prediction on the transitional regime from Newtonian to modified dynamics. This is especially unfortunate given the fact that the transitional regime has been explored by observations: the empirical \emph{mass discrepancy--acceleration (MDA) relation} \citep{mcgaugh2004} shows that the ratio $\Mt/\Mo$ (the mass discrepancy) is a characteristic function of the accelerations $\ac$ and $\gn$; here $\Mt$ is the total dynamical mass given by $\ac=G\Mt/r^2$. If a prediction for $\mu(x)$ were available, it could be tested by comparison to the empirical MDA relation in a straightforward manner.

Recently, I proposed a scheme for gravitational interaction on galactic scales (the ``graviton picture'') which is based on the ad-hoc assumption that gravity is mediated by virtual massive gravitons that obey certain reasonable rules of interaction \citep{trippe2013}. The ``graviton picture'' predicts a theoretical MDA relation which is in good agreement with observations; it comprises expressions for limiting cases that agree with those of MOND (and likewise agree with observations). The present work follows up on, and amends, \citet{trippe2013}. I realized only after publication of \citet{trippe2013} an additional consequence of my toy model of gravity introduced there: A comparison of classical MOND and ``graviton picture'' shows that MOND naturally derives from the ``graviton picture'' at least for the case of non-relativistic, highly symmetric dynamical systems. This comparison is the subject of the present work.

\section{ANALYSIS}

\begin{figure*}[!t]
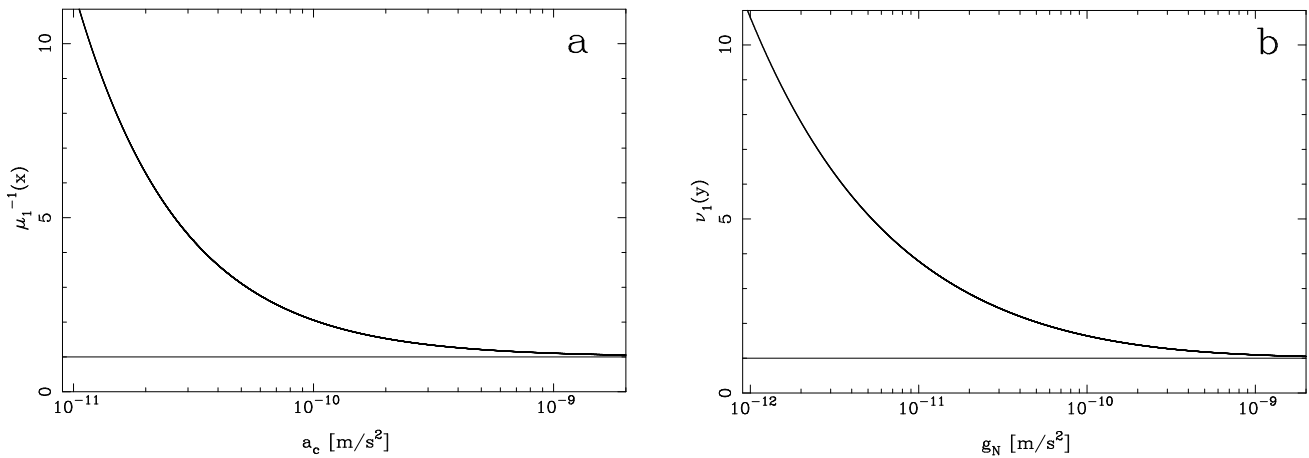

\plotfiddle{md-ac-mu-theo.eps}{55mm}{-90}{35}{35}{-259}{190}
\plotfiddle{md-gn-nu-theo.eps}{0mm}{-90}{35}{35}{-1}{211}
\caption{The transition functions derived from comparison of MOND and ``graviton picture'', assuming $\am=1.1\times10^{-10}$\,m\,s$^{-2}$. {\bf a.} The inverse of the function $\mu_1(x)$ vs. $\ac$. {\bf b.} The function $\nu_1(y)$ vs. $\gn$. Please note the logarithmic--linear axis scales. These diagrams should be compared to Fig.~10 of \citet{famaey2012} and Fig.~11 of \citet{kroupa2012}.}
\label{fig_mu}
\end{figure*}

\subsection{Transition Functions in MOND}

\noindent
Equation~\ref{eq_mu} provides the defining properties of MOND; the transition function $\mu(x)$ is constrained (1) by limiting cases that have to be consistent with observations plus (2) the condition that $x\mu(x)$ increases monotonically with $x$ \citep{famaey2012}. These conditions are fulfilled (e.g., \citealt{milgrom1983a,famaey2005,famaey2012}) by the set of functions

\begin{equation}
\label{eq_mu-n}
\mu_n(x) = \frac{x}{\left(1+x^n\right)^{1/n}} ~ ; ~~~ n=1,2,3,... ~ .
\end{equation}

\noindent
Alternatively, one may re-write Eq.~\ref{eq_mu} as

\begin{equation}
\label{eq_nu}
\frac{\ac}{\gn} = \nu(y)
\end{equation}

\noindent
where $y=\gn/\am$; $\nu(y)\rightarrow1$ for $y\gg1$ and $\nu(y)\rightarrow y^{-1/2}$ for $y\ll1$. For reasons analogous to those for the case of $\mu(x)$, a set of valid transition functions is given by \citep{famaey2012}

\begin{equation}
\label{eq_nu-n}
\nu_n(y) = \left\{\frac{1}{2}\left[ 1 + \left(1 + \frac{4}{y^n}\right)^{1/2} \right]\right\}^{1/n} .
\end{equation}

\noindent
Comparison to the empirical MDA relation suggests $n=1$ or $n=2$ \citep{kroupa2012}. Nevertheless, none of these scaling relations follows from first principles: technically, arbitrary alternative transition functions can be constructed from comparison to the data.

\subsection{The ``Graviton Picture'' \label{ssect_gravitons}}

\noindent
The ``graviton picture'' \citep{trippe2013} employs the ad-hoc assumption that gravity is mediated by virtual gravitons with non-zero mass that obey certain reasonable rules of interaction. This leads to the formation of a ``graviton halo''  with a mass density profile $\rho\propto r^{-2}$ around a baryonic source mass $\Mo$. The total dynamical mass $\Mt$ scales with centripetal acceleration $\ac$, providing a theoretical MDA relation

\begin{equation}
\label{eq_mda}
\frac{\Mt}{\Mo} = 1 + 8\pi\frac{\ao}{\ac}
\end{equation}

\noindent
where $\ao$ is a constant of the dimension of an acceleration. The limiting case $\ac\gg8\pi\ao$ corresponds to the usual Newtonian dynamics. The limiting case $\ac\ll8\pi\ao$ leads to $\vc^4\approx8\pi\,G\Mo\,\ao=const.$; from comparison to the corresponding MOND result we find $\am\equiv8\pi\ao$. 

Else than MOND, the ``graviton picture'' comprises a scaling law for the transitional regime from Newtonian to modified dynamics (Eq.~\ref{eq_mda}) \emph{a priori}. In analogy to Eq.~\ref{eq_mu} we can define a transition function

\begin{equation}
\label{eq_xi}
\xi(x) = \frac{\gn}{\ac} = \frac{\Mo}{\Mt} = \left[ 1 + \frac{1}{x} \right]^{-1}
\end{equation}

\noindent
with $x=\ac/\am$ as before. This results in

\begin{equation}
\label{eq_xi-mu}
\xi(x) = \frac{x}{1 + x} = \mu_1(x) ~ .
\end{equation}

\noindent
As we see, the transition function $\xi(x)$ provided by the ``graviton picture'' is identical to the MOND transition function $\mu_n(x)$ for $n=1$.

Furthermore, we can define an inverse transition function in analogy to Eq.~\ref{eq_nu} like

\begin{equation}
\label{eq_zeta-def}
\zeta = \frac{\Mt}{\Mo} = \frac{\ac}{\gn} ~~ \longrightarrow ~~ y = \frac{\gn}{\am} = \frac{x}{\zeta} ~ .
\end{equation}

\noindent
Via Eq.~\ref{eq_mda} this leads to the quadratic equation

\begin{equation}
\label{eq_zeta-eq}
\zeta^2 - \zeta - \frac{1}{y} = 0 ~ ;
\end{equation}

\noindent
solving this expression for $\zeta$ and ignoring the unphysical negative root, we find

\begin{equation}
\label{eq_zeta}
\zeta(y) = \frac{1}{2} \left[ 1 + \left( 1 + \frac{4}{y} \right)^{1/2} \right] = \nu_1(y) ~ .
\end{equation}

\noindent
In this case, the transition function $\zeta(y)$ provided by the ``graviton picture'' is identical to the MOND transition function $\nu_n(y)$ for $n=1$. I illustrate the functions $\xi^{-1}(x)=\mu_1^{-1}(x)$ and $\zeta(y)=\nu_1(y)$ in Fig.~\ref{fig_mu}; the diagrams should be compared to Fig.~10 of \citet{famaey2012} and Fig.~11 of \citet{kroupa2012}.

\section{DISCUSSION}

\begin{quote}
\small
{\bf ``}\,It is principally the elegance of general relativity theory and its success in solar system tests that lead us to the bold extrapolation that the gravitational acceleration has the form $GM/r^2$ on scales $10^{21}-10^{26}$\,cm that are relevant for the solar neighborhood, galaxies, clusters of galaxies, and superclusters.\,{\bf ''}

--- \citet{binney1987}, Ch.\,10.4, p.\,635
\end{quote}

\noindent
Historically, the missing mass problem has usually been approached by postulating non-luminous and non-baryonic \emph{dark matter} \citep{ostriker1973,einasto1974}, eventually evolving into the $\Lambda$CDM standard model of cosmology (e.g., \citealt{bahcall1999}). In recent years, it has become clear that this approach is incomplete. The assumption of dark matter distributed within and around galaxies is partially incompatible with observations of structure and dynamics of galaxies and groups of galaxies (see \citealt{kroupa2012} for a recent review). Standard cosmology seems unable to predict fundamental relations of galactic dynamics like the Faber-Jackson and Tully-Fisher relations \citep{faber1976,tully1977}, the MDA relation, or the surface brightness--acceleration relation. More generally, it has been found that galactic dynamics is intimately linked with a universal characteristic acceleration which can be identified with Milgrom's constant $\am\approx10^{-10}$\,m\,s$^{-2}$ (e.g., \citealt{famaey2012}).

The dynamics of galaxies and groups of galaxies can be understood in the frame of theories of modified gravity and/or inertia based on acceleration scales -- a discovery eventually leading to MOND (e.g., \citealt{milgrom1983a,bekenstein1984,sanders2002,ferreira2009}). On the one hand, MOND has been remarkably successful in describing galactic dynamics while circumventing the problems that plague $\Lambda$CDM cosmology (cf. Sect.~\ref{sect_intro}; also \citealt{ferreira2009,famaey2012,kroupa2012}). On the other hand, MOND is obviously incomplete: it does not derive from a physical mechanism but rather empirically from the boundary conditions provided by observations. In addition, it does not comprise a prediction for the transitional regime from Newtonian to modified dynamics; appropriate scaling laws have to be constructed from comparison to observations.

The ``graviton picture'' of gravitation \citep{trippe2013} starts off from a physical mechanism: the (ad-hoc) assumption that gravity is mediated by virtual massive gravitons that obey certain rules of interaction. From this it is possible to derive the scaling law given by Eq.~\ref{eq_mda} which is in good agreement with observations. As shown in Sect.~\ref{ssect_gravitons}, this scaling law actually comprises the empirical scaling laws employed in the frame of MOND -- at least for non-relativistic dynamical systems and circular orbit geometries.

The ``graviton picture'' as well as MOND each comprise a characteristic acceleration, either $\am$ or $\ao$, with $\am=8\pi\ao$. These characteristic accelerations are free parameters. Empirically, $\am\approx c\,H_0/2\pi$ with $c$ denoting the speed of light and $H_0$ denoting Hubble's constant (e.g., \citealt{famaey2012}). This might suggest a physical connection between galactic dynamics and cosmology; however, as yet this relation is entirely empirical and has not been derived from first principles.

Evidently, the new feature I add to the discussion of MONDian scaling laws of gravitation is the theoretical MDA relation (Eq.~\ref{eq_mda}). This relation implies a proportionality of total dynamical mass $\Mt$ and luminous baryonic mass $\Mo$ -- a property expected for modified laws of gravity but sharply distinct from standard dark matter models where dark and luminous mass are distributed independently. Such a tight relation between luminous and baryonic mass is observed in galactic rotation curves (known as ``Renzo's rule''; \citealt{sancisi2004}) and in the ``Train Wreck'' (A520) cluster of galaxies \citep{jee2012}. A possible counter-example is provided by the ``Bullet Cluster'' of galaxies (1E0657-56); here observations indicate a spatial separation of dark and luminous mass but \emph{also} find the cluster kinematics to be inconsistent with $\Lambda$CDM cosmology \citep{lee2010} -- making the overall analysis of the ``Bullet Cluster'' rather inconclusive actually. A more detailed discussion of various observational tests of the ``graviton picture'' is provided in \citet{trippe2013}.

Regarding the combined evidence, I conclude that (classical) MOND naturally derives from the ``graviton picture''. Eventually, this suggests that the ``graviton picture'' -- despite its toy-model character \citep{trippe2013} -- provides a valid candidate for the physical mechanism behind MOND and gravity on galactic scales in general -- at least to first order.

\section{CONCLUSIONS}

\noindent
I compare Modified Newtonian Dynamics with the ``graviton picture'' of gravitation which I proposed recently, and arrive at the following principal conclusions:

\begin{enumerate}

\item  MOND and the ``graviton picture'' find identical expressions for dynamics in the limits of strong and weak centripetal accelerations. Their parameters, $\am$ and $\ao$ respectively, are related like $\am\equiv8\pi\ao$; empirically, $\am\approx10^{-10}$\,m\,s$^{-2}$.

\item  The empirical scaling laws commonly used in MOND, which are needed to quantify the transitional regime between Newtonian and modified dynamics, are contained in the ``graviton picture'' a priori.

\item  MOND derives from the ``graviton picture'' -- at least for non-relativistic, highly symmetric dynamical systems.

\end{enumerate}

\noindent
These conclusions suggest that -- at least to first order -- the ``graviton picture'' provides a valid candidate for the physical mechanism behind MOND and gravity on galactic scales in general.

\acknowledgments{\noindent\small I am grateful to {\sc Junghwan Oh}, {\sc Taeseok Lee}, {\sc Jae-Young Kim}, and {\sc Jong-Ho Park} (all at SNU) for valuable technical support. This work made use of the software package {\sc dpuser}\footnote{\tt http://www.mpe.mpg.de/$\sim$ott/dpuser/dpuser.html} developed and maintained by {\sc Thomas Ott} at MPE Garching. I acknowledge financial support from the Korean National Research Foundation (NRF) via Basic Research Grant 2012R1A1A2041387. Last but not least, I am grateful to an anonymous referee whose comments helped to improve the quality of this paper.}

\end{document}